# Field dependent permittivity of composite materials containing ferromagnetic wires

D. P. Makhnovskiy and L. V. Panina

*Department of Communication, Electronics and Electrical Engineering,*

*University of Plymouth, Drake Circus, Plymouth, Devon PL4 8AA, United Kingdom.*

**Abstract** A new type of a composite material is proposed, the microwave permittivity of which changes under the effect of a dc magnetic field applied to the whole composite sample. The composite consists of short ferromagnetic wires embedded into a dielectric matrix. A strong field dependence of the permittivity is seen in the vicinity of the antenna resonance, where the dispersion behaviour can experience a transformation from a resonant spectrum to a relaxation one under the effect of the field. This permittivity behaviour owes to a high sensitivity of the ac surface impedance of a ferromagnetic wire to a magnetic field, known as the magneto-impedance (MI) effect. If the resonance-like dispersion behaviour is realised, the real part of the effective permittivity can be made negative past the resonance for wire inclusion concentrations well below the percolation threshold. Applying a magnetic field, the negative peak continuously decreases as the dispersion tends to become of a relaxation type. The effective permittivity is analysed within a one-particle approximation, by considering a wire piece as an independent scatterer and solving the scattering problem with the impedance boundary condition. A magnetic field is assumed to be applied in parallel to the wire. A new integro-differential equation for the current distribution in a wire is obtained, which is valid for the surface impedance matrix of a general form. This work demonstrates a possibility of using the MI effect to design field-controlled composites and band-gap structures.

## I. Introduction

Numerous studies have been dedicated to the electrodynamics of metal-dielectric composites, in particular, to the role of microstructure in determining the effective permittivity and permeability.[1] In the present work, a new type of the composite material is advanced, for which the effective microwave permittivity $\varepsilon_{eff}$ can be controlled by a static magnetic field $H_{ex}$. Short pieces of ferromagnetic microwires[2-4] are proposed as filling inclusions. They interact with the electromagnetic radiation similar to microantennas. Then, the wire length $l$ and dielectric matrix permittivity $\varepsilon$ define the operating frequency range with the characteristic frequency related to the antenna resonance. At certain conditions, the microwave permeability matrix of the wire, and hence its surface impedance, are very sensitive to $H_{ex}$. In the vicinity of the antenna resonance, the variations in the surface impedance result in a considerable change in the current distribution at the wire and, consequently, in the induced dipole moment of an elementary wire-scatterer. A large change in the high-frequency surface impedance of ferromagnetic materials subjected to a dc magnetic field is known as the magneto-impedance (MI) effect.[5] Up to the present, the MI effect has been used for highly sensitive magnetic sensors, however, it appears that the applications of this effect can be much wider. Typically, MI sensors are designed for MHz frequencies,[6] which is dictated by the related electronics. On the other hand, the field sensitivity of the surface impedance in wires with a circumferential anisotropy remains very high even at the GHz range. Thus, the MI effect can be useful to design microwave composites with tuneable properties and tuneable band-gap structures.[7]

Composite materials containing elongated conducting inclusions – finite length wire or arrays of infinitely long wires – present a considerable interest since their dielectric response can exhibit various dispersive behaviours. Along with this, the nominal values of the effective permittivity are very large even for small concentrations. So far, the materials with nonmagnetic inclusions embedded into a dielectric matrix with permittivity $\varepsilon$ have been considered. Finite-wire inclusions behave as electric dipole scatterers. In this case, the dispersion of the effective permittivity can range from a relaxation type to a resonance one, depending on the inclusion conductivity and dimension.[8-10] The resonant spectrum of $\varepsilon_{eff}$ occurs near the antenna resonance for an individual wire-inclusion. The real part of $\varepsilon_{eff}$ can take negative values past the resonance, which is very important for a recent trend to create materials with a negative refraction index.[11-13] This characteristic behaviour of the effective permittivity was experimentally confirmed for diluted composites with carbon fibres having a low conductivity (relaxation spectrum) and highly conductive aluminium–coated fibres (resonant spectrum).[9]

Here we demonstrate that the physical quantity controlling the dispersion characteristics of $\varepsilon_{eff}$ in diluted wire-composites is the surface impedance, which involves both the conductivity and permeability of inclusions. Therefore, in the case of ferromagnetic wires the effective permittivity may depend on a static magnetic field via the corresponding dependence of the ac permeability matrix. The theory developed is based on solving the scattering problem for a wire with the impedance boundary conditions. The new integro-differential equation for the current distribution in a wire is obtained, which is valid for the surface impedance matrix of a general form. The electric polarisability of an inclusion is represented by the integral over the longitudinal current density and is proven to be very sensitive to the impedance changes near the antenna resonance.

In Ref. 10, the current distribution at a conducting stick was determined from an approximate differential equation of a transmission line type. Our approach has a number of advantages. It gives a rigorous mathematical algorithm as an expansion in serious of $1/2\ln(l/a)$ ($a$ is the wire radius) valid for any frequency and the surface impedance of a general matrix form. There is no need to introduce the effective distributed parameters such as a specific capacitance and an inductance which have to be determined separately. The most important is that the method accounts correctly for the radiation effects and can be generalised to the case of interacting inclusions or inclusions interacting with boundaries and interfaces. Then, it can be useful to calculate the effective permittivity of composites containing periodically spaced wires, without invoking the effective medium theory. For thin composite sheets, the effective permittivity depends on the thickness due to the depolarisation effect from the boundaries.[14,15] As a result, the dispersion region of $\varepsilon_{eff}$ is shifted to higher frequencies.

Another interesting material is a composite system made of an array of infinitely long conductive wires.[16,17] This material has characteristic features of a metallic response to radiation, but in the GHz range. Contrary to the composite with short inclusions, the electromagnetic field is applied locally to a certain portion of the material excluding the ends of wires. In this case the current distribution in the wire can be neglected. The most interesting results are obtained for the wave polarisation where the electric field is along the wires. Such wire-mesh systems model the response of a diluted plasma,[1,16] giving a negative permittivity $\varepsilon_{eff}(\omega)$ below the normalised plasma frequency $\tilde{\omega}_p = \omega_p / \sqrt{\varepsilon}$ somewhere in the gigahertz range: $\varepsilon_{eff}(\omega) = \varepsilon - \omega_p^2 / \omega^2$, where $\varepsilon$ is the matrix permittivity and $\omega_p$ is the "plasma frequency". In a general case, when the skin effect is not very strong, the plasma frequency depends on the wire impedance.[1] Therefore, the effective permittivity of wire-mesh materials can be also controlled by a magnetic field, as will be considered elsewhere.

## II. General approach to the effective permittivity of wire-composites

In this Section we consider general properties of the effective permittivity of composites with finite-wire inclusions. From the above discussion it follows that the dispersion of $\varepsilon_{eff}$ has different origins for "short" and "long" inclusions. In the first case, the composites demonstrate the Lorentz dipole dispersion, whereas the second type of materials is characterised by the Drude dispersion of free-electron gas.

The Lorentz model of dispersion is applicable to insulator materials. The composite with short inclusions is similar in many respects to an isolator since the wire-inclusions play a role of “atoms” (elementary dipole scatterers), which are polarised with an ac electric field. The local electrical field $e_{loc}\exp(-i\omega t)$ induces the current with a linear density $j(x)\exp(-i\omega t)$ distributed along the inclusion length. The electric dipole moment $D$ and the dielectric polarisability $\alpha$ of the inclusion are calculated using the continuity equation $\partial j(x)/\partial x = i\omega\rho(x)$ and integrating by parts with boundary conditions $j(\pm l/2)\equiv 0$ ($\rho$ is the charge density per unit length):

$$D = \frac{i}{\omega}\int_{-l/2}^{l/2} j(x)dx\,, \quad \alpha = D/(V e_{loc})\,, \tag{1}$$

where $V$ is the inclusion volume. As it will be shown later, the density $j(x)$ of a linear current can be approximated by a linear differential equation of the second order with the boundary conditions $j(\pm l/2)\equiv 0$ and involving a certain damping caused by radiation and internal resistive and magnetic losses. Thus, as in the case of a Lorentz oscillator the polarisability $a$ has the following form:[18]

$$\alpha = \sum_n \frac{A_n}{(\omega_{res,n}^2 - \omega^2) - i\Gamma_n\omega}\,, \tag{2}$$

where the summation is carried out over all antenna resonance frequencies $\omega_{res,n} = 2\pi c/\lambda_{res,n}$ in increasing order, $\lambda_{res,n} = 2l\sqrt{\varepsilon}/(2n-1)$[19] is the resonance wavelengths, $\varepsilon$ is the matrix permittivity, $A_n$ is the amplitude constants, $\Gamma_n$ is the dumping parameter. The first resonance $n=1$ with the lowest frequency has a maximum amplitude $A_1$ and gives the main contribution to the polarisability. Each $\Gamma_n$ can be decomposed into two parts $\Gamma_n^{rad}$ and $\Gamma_n^{mr}(H_{ex})$ related to the radiation and internal (magnetic and resistive) losses, respectively. The damping parameter $\Gamma_n^{mr}$ involving magnetic losses may depend on an external magnetic field $H_{ex}$. Thus, in the vicinity of the antenna resonance the polarisability $\alpha$ will depend on $H_{ex}$ if the condition $|\Gamma_n^{mr}| \sim |\Gamma_n^{rad}|$ is held.

The bulk polarisation P of the composite is of the form: $\mathrm{P} = < e_{loc} > p\alpha = e_0 \vartheta_{eff}$, where $< e_{loc} >$ is the averaged local field, $p$ is the volume concentration of the inclusions, $\mathbf{e}_0$ is the external electrical field, and $J_{eff}$ is the effective bulk susceptibility. Although the wire length is comparable with the wavelength, it is still possible to introduce the susceptibility $J_{eff}$ since the scattered electromagnetic field has a dipole character at large distances from the composite. To relate the polarisability $\alpha$ to the effective bulk susceptibility $J_{eff}$, the relation between $e_{loc}$ and $e_0$ has to be established. For small inclusion concentrations $p << p_c$, where $p_c$ is the percolation threshold, it is reasonable to assume that $< e_{loc} > \approx e_0$, which leads to:

$$\varepsilon_{eff} \approx \varepsilon + 4\pi p < \alpha >, \tag{3}$$

where $< \alpha >$ is the polarisability averaged over the inclusion orientations. In the limit $p << p_c$, the difference between the local and mean fields can be taken into account using the Lorentz approach[18] and its generalisations.[20] A consistent effective medium theory for the considered composites valid for any concentration employs the concept of the scale-dependent local-field permittivity.[10,9,15] In this work we restrict ourselves to the case of non-interacting inclusions since this model provides all the essential features of the dispersion of $\varepsilon_{eff}$ and its magnetic field dependence in diluted composites.

**III. Antenna approximation and impedance boundary conditions**

Within the framework of a single particle approximation the scattering problem for a thin conductor has to be solved. Considering the electromagnetic response from a thin conductor, the induced current in it can be replaced with the effective linear current flowing along the axis and having only the axial distribution. This approach is known as the antenna approximation.[21] It is important to re-examine the conditions when the antenna approximation is valid.

Let us consider a thin conductor irradiated by an electromagnetic field. The wavelength $\lambda$ and the conductor length $l$ are assumed to be much larger than the conductor cross size $2a$: $2a << \lambda$ and $2a << l$. The incident electromagnetic wave is supposed to be of a plane type. In this case, the external electric field does not induce a circular current, and the external magnetic field does not give a contribution to a circular magnetic field on the conductor surface. First we consider that the incident wave has a longitudinal electric field $\bar{e}_{x0}$ at the surface of a nonmagnetic conductor ($x$ is the coordinate along the conductor). In this case, the induced current is longitudinal, which determines the scattered electromagnetic field having longitudinal electric $\bar{e}_x$ and circular magnetic $\bar{h}_\varphi$ components on the conductor surface (cylindrical conductor is considered, $\varphi$ is the azimuthal coordinate). The same polarisation $(\bar{e}_x, \bar{h}_\varphi)$ can be induced by a linear current with the volume density $j(x)\delta_s$ flowing along the axis, where $x$ is a point on the axis and $\delta_s$ is two dimensional Dirac's function. Further the function $j(x)$ will be referred to as "linear density" or "density". Thus, the linear longitudinal current plays a role of an effective current producing the surface field of the required polarisation $(\bar{e}_x, \bar{h}_\varphi)$ and intensity. If the incident electromagnetic field contains a longitudinal magnetic component $\bar{h}_{x0}$, a circular electric field $\bar{e}_\varphi$ will be induced in the conductor. In this case, a longitudinal linear current does not provide the total polarisation of the scattered field. In a general case of a magnetic conductor the field $\bar{h}_{x0}$ will induce $\bar{e}_x$ and the field $\bar{e}_{x0}$ will induce $\bar{h}_\varphi$. However, the total scattered field can be decomposed into two basic waves with polarisations: $(\bar{e}_x, \bar{h}_\varphi)$ and $(\bar{e}_\varphi, \bar{h}_x)$, where $(\bar{e}_x, \bar{h}_\varphi)$ is determined by the linear current. The other polarisation $(\bar{e}_\varphi, \bar{h}_x)$ can be calculated directly from the impedance boundary condition, which represents a linear relationship between $\bar{e}$ and $\bar{h}$ (see below). The polarisation effects arising due to the radial electric and magnetic fields can be neglected in this case. Thus, the concept of the linear effective current describes correctly the scattered field at any polarisation of the incident wave.

In the antenna approximation, the response from a thin inclusion irradiated by an electromagnetic field is fully determined by the external scattering problem with the boundary conditions at the inclusion, which is set via the surface impedance matrix $\hat{\varsigma}$:[22]

$$\bar{\mathbf{E}}_t = \hat{V}(\bar{\mathbf{H}}_t \times \mathbf{n}), \tag{4}$$

where $\mathbf{n}$ is the unit normal vector directed inside the conductor, $\bar{\mathbf{E}}_t$ and $\bar{\mathbf{H}}_t$ are the tangential vectors of the total electric and magnetic fields at the conductor surface, which include both the scattered and external fields. In the case of ideal conductor (conductivity $\sigma = \infty$) condition (4) nulls: $\bar{\mathbf{E}}_t \equiv 0$, which is typically used in the antenna problems. Boundary condition (4) is convenient to write in the local cylindrical co-ordinate system $(x, j, r)$ related to the conductor:

$$\begin{aligned} \bar{E}_x &= \varsigma_{xx}\bar{H}_\varphi - \varsigma_{x\varphi}\bar{H}_x \\ \bar{E}_\varphi &= \varsigma_{\varphi x}\bar{H}_\varphi - \varsigma_{\varphi\varphi}\bar{H}_x \end{aligned}, \tag{5}$$

The matrix $\hat{\varsigma}$ was found in Ref. 23 for a ferromagnetic wire with an arbitrary type of the magnetic anisotropy for any frequency. In a nonmagnetic conductor off-diagonal terms $\varsigma_{x\varphi} = \varsigma_{\varphi x} \equiv 0$.

Within the antenna approximation, the field $\bar{H}_\varphi(x)$ contains only the circular field $\bar{h}_\varphi(x)$ induced by a current with linear density $j(x)$. On the contrary, the longitudinal field $\bar{H}_x$ is entirely defined by the excitation field. The scattered field $\bar{h}_\varphi(x)$ will be found together with the antenna equation in the next Section.

## IV. Antenna equation with the impedance boundary condition

Now we are in a position to obtain the basic integro-differential equation for the current density $j(x)$ using impedance boundary condition (5). The time dependence is taken as $\exp(-i\omega t)$, where $\omega = 2\pi f$ and $f$ is the frequency of the electromagnetic field. Gaussian units are used through out the paper.

Let us introduce the vector $\mathbf{A}$ and scalar $\varphi$ potentials:

$$\mathbf{h} = \frac{4\pi}{c}\operatorname{rot}\mathbf{A}, \qquad \mathbf{e} = -\operatorname{grad}\varphi - \frac{4\pi\mu}{c^2}\frac{\partial \mathbf{A}}{\partial t}, \tag{6}$$

where $c$ is the velocity of light, $\varepsilon$ and $\mu$ are the dielectric and magnetic constants outside the conductor. The Lorentz gauge is accepted for the potentials: $\varepsilon\partial\varphi/\partial t + 4\pi \operatorname{div}\mathbf{A} = 0$. The electrical field $\mathbf{e}$ can be expressed through the vector potential in a frequency representation:

$$\mathbf{e} = \frac{4\pi i\omega\mu}{c^2}\mathbf{A} - \frac{4\pi}{i\omega\varepsilon}\operatorname{grad}\operatorname{div}\mathbf{A}. \tag{7}$$

For $\mathbf{A}$ we obtain the Helmholtz equation:

$$\Delta\mathbf{A} + k^2\mathbf{A} = \mathbf{j}, \tag{8}$$

where $k = (\omega/c)\sqrt{\varepsilon\mu}$ is the wave number, $\mathbf{j}$ is the vector of the current density $j(x)$.

The solution of Eq. (8) can be written in the form of the convolution of $\mathbf{j}(\mathbf{x})$ with the Green function $G(r)$ of the Helmholtz operator, the formalism of which has been developed in Ref. 15:

$$\mathbf{A}(\mathbf{r}_0) = (G * \mathbf{j}) = \int_V \mathbf{j}(\mathbf{x}) G(r) dV_{\mathbf{x}}, \qquad G(r) = \frac{\exp(ikr)}{4pr}, \tag{9}$$

where the integration is carried out over the total volume $V$ containing $\mathbf{j}$. In Eq. (9) $\mathbf{r}_0$ is the co-ordinate of the point where $\mathbf{A}$ is calculated (observation point), $\mathbf{x}$ is the vector directed to the integration point, and $r = |\mathbf{r}_0 - \mathbf{x}|$.

From Eqs. (6) and (9) we obtain the representation for the magnetic field induced by a linear current :

$$\mathbf{h}(\mathbf{r}_0) = \frac{1}{c} \int_V \frac{(1 - ikr)\exp(ikr)}{r^3} (\mathbf{j}(\mathbf{x}) \times \mathbf{r})\, dV_{\mathbf{x}}, \tag{10}$$

where $\mathbf{r} = \mathbf{r}_0 - \mathbf{x}$. In the case of a linear current, when $\mathbf{j}$ is taken at the contour $L$, the field projection on the unit vector $\upsilon$ reduces to a kind of a contour integral:

$$h(\mathbf{r}_0) = \frac{1}{c} \int_L \frac{(1 - ikr)\exp(ikr)}{r^3}\, j(s) \xi(s, \mathbf{r}) ds, \tag{11}$$

where $\mathbf{j} = j(s)\tau_s$, $\tau_s$ is the tangential vector along $L$ taken at the integration point $s$, $x(s, \mathbf{r}) = ((t_s \times \mathbf{r}) \cdot u)$ is the scalar product designating the projection of $(t_s \times \mathbf{r})$ on the unit vector $\upsilon$. For a cylindrical symmetry, a circular magnetic field is equal to a projection of $\bar{\mathbf{h}}(\mathbf{r}_0)$ on the direction $(t_s \times \mathbf{r})$:

$$\bar{h}_j(x, a) = \frac{a}{c} \int_{-l/2}^{l/2} \frac{(1 - ikr)\exp(ikr)}{r^3}\, j(s) ds, \tag{12}$$

where $r = \sqrt{(x - s)^2 + a^2}$. In (12) the equality $|(t_s \times \mathbf{r})| = a$ was used. Contrary to the static case $(w = 0)$ where $\bar{h}_\varphi = 2I/ac$ and $I$ is the total current, Eq. (12) takes into account the retarding effects. Note that integral (12) has extremely fast convergence, therefore, the field $\bar{h}_\varphi$ appears to be almost local even for very high frequencies.

The component $A_x$ of the vector potential **A** describes the scattered field from a straight piece of a thin conductor. Using Eqs. (7) and (9), the longitudinal scattered field $e_x(x,y,z)$ can be expressed in terms of an integro-differential operator, where the convolution is carried out along the longitudinal co-ordinate $x$:

$$e_x(x,y,z) = -\frac{4\pi}{i\omega\varepsilon}\left[\frac{\partial^2}{\partial x^2}(G*j) + k^2(G*j)\right], \qquad (13)$$

$$(G*j) = \int_{-l/2}^{l/2} j(s)G(r)ds, \qquad G(r) = \exp(ikr)/4p\,r,$$

where $r = \sqrt{(x-s)^2 + y^2 + z^2}$. On the conductor surface, it is necessary to put $r = \sqrt{(x-s)^2 + a^2}$. Using the impedance boundary condition (5) and Eq. (13) we obtain the integro-differential equation for the current density $j(x)$:

$$\bar{E}_x \equiv -\frac{4\pi}{i\omega\varepsilon}\left[\frac{\partial^2}{\partial x^2}(G*j) + k^2(G*j)\right] + \bar{e}_{0x}(x) = \varsigma_{xx}\bar{h}_\varphi(x) - \varsigma_{x\varphi}\bar{h}_{0x}(x), \qquad (14)$$

where $\bar{E}_x \equiv \bar{e}_x + \bar{e}_{0x}$ is the total longitudinal electric field on the conductor surface, $\bar{e}_x$ is the scattered electrical field, $\bar{e}_{0x}$ and $\bar{h}_{0x}$ are the external electrical and magnetic fields. The components $\varsigma_{xx}$ and $\varsigma_{x\varphi}$ can also be functions of $x$, but this case is not considered here.

The surface field $\bar{h}_\varphi$ is convenient to write in terms of convolution with $j(x)$:

$$\bar{h}_\varphi(x,a) = \frac{2}{ac}(G_\varphi * j) = \frac{2}{ac}\int_{-l/2}^{l/2} j(s)G_\varphi(r)ds, \qquad (15)$$

where

$$G_\varphi(r) = \frac{a^2(1-ikr)\exp(ikr)}{2r^3}.$$

Finely, we obtain the basic integro-differential equation for $j(x)$:

$$\frac{\partial^2}{\partial x^2}(G*j)+k^2(G*j)=\frac{i\omega\varepsilon}{4\pi}\bar{e}_{0x}(x)-\frac{i\omega\varepsilon\varsigma_{xx}}{2\pi a c}(G_\varphi*j)+\frac{i\omega\varepsilon\varsigma_{x\varphi}}{4\pi}\bar{h}_{0x}(x). \qquad (16)$$

Equation (16) has to be completed imposing the boundary conditions at the ends of the conductor:

$$j(-l/2)=j(l/2)\equiv 0 \qquad (17)$$

The antenna equation for a nonmagnetic conductor with ideal conductivity $s=\infty$ embedded into a dielectric layer has been obtained in Ref. 15, which accounts for only the radiation losses. *New integro-differential equation (16) involves general losses including both radiation and internal losses (resistive losses and magnetic relaxation). The internal losses appear via impedance* $\varsigma_{xx}$ *and the convolution* $(G_\varphi*j)$, *whereas the imaginary part of* $(G*j)$ *determines the radiation losses.*

Along with this, there is an additional term in the right part of Eq. (16), related with the off-diagonal component $\varsigma_{x\varphi}$. Thus, the ferromagnetic conductor can be excited not only by a longitudinal electric field, but also by a longitudinal magnetic field.[23]

As it follows from Eqs. (13) and (15), the real functions $\mathrm{Re}(G)$ and $\mathrm{Re}(G_\varphi)$, considered at the conductor surface, have a sharp peak at $r=a$. Thus, $\mathrm{Re}(G)$ and $\mathrm{Re}(G_\varphi)$ give the main contribution to Eq. (16): $|(\mathrm{Im}(G)*j)|<<|(\mathrm{Re}(G)*j)|$ and $|(\mathrm{Im}(G_\varphi)*j)|<<|(\mathrm{Re}(G_\varphi)*j)|$. However, the convolutions with the imaginary parts are important in the vicinity of the resonance and can be taken into account by an iteration procedure, which is described in Appendix. For the calculation of convolutions with functions $\mathrm{Re}(G)$ and $\mathrm{Re}(G_\varphi)$ it is possible to use an approximate method:[24,14,15]

$$(\mathrm{Re}(G)*j)\approx j(x)\int_{-l/2}^{l/2}\mathrm{Re}(G(r))ds=j(x)Q,$$

$$Q=\int_{-l/2}^{l/2}\mathrm{Re}(G(r))ds \propto \frac{1}{4p}\int_{-l/2}^{l/2}\frac{ds}{\sqrt{s^2+a^2}} \sim \frac{\ln(l/a)}{2p}, \tag{18}$$

$$(\mathrm{Re}(G_\varphi)*j)\approx j(x)\int_{-l/2}^{l/2}\mathrm{Re}(G_\varphi(r))ds = j(x)Q_\varphi ,$$

$$Q_\varphi=\int_{-l/2}^{l/2}\mathrm{Re}(G_\varphi(r))ds \propto \frac{a^2}{2}\int_{-l/2}^{l/2}\frac{ds}{(s^2+a^2)^{3/2}}+\frac{a^2k^2}{2}\int_{-l/2}^{l/2}\frac{ds}{\sqrt{s^2+a^2}} \propto (1+a^2k^2\ln(l/a)) \sim 1,$$

where $r=\sqrt{(x-s)^2+a^2}$ , $Q$ and $Q_\varphi$ are the positive form-factors. For the estimation of $Q_\varphi$ it was taken into account that $ak<<1$ in the antenna approximation.

From Eq. (16) and inequalities $|(\mathrm{Im}(G)*j)|<<|(\mathrm{Re}(G)*j)|$ and $|(\mathrm{Im}(G_\varphi)*j)|<<|(\mathrm{Re}(G_\varphi)*j)|$ we obtain differential equation for the zero approximation $j_0(x)$ where the radiation losses are neglected:

$$\frac{\partial^2}{\partial x^2}j_0(x)+\left(\frac{\omega}{c}\right)^2\varepsilon\mu\left(1+\frac{ic\,\varsigma_{xx}}{2\pi a\omega\mu}\frac{Q_\varphi}{Q}\right)j_0(x)\approx\frac{i\omega\varepsilon}{4\pi Q}\left[\bar{e}_{0x}(x)+\varsigma_{x\varphi}\bar{h}_{0x}(x)\right]. \tag{19}$$

As it follows from Eq. (19), the implementation of the impedance boundary condition leads to the renormalisation of the wave number which becomes:

$$\tilde{k}=\frac{\omega}{c}\sqrt{\varepsilon\mu}\left(1+\frac{ic\varsigma_{xx}}{2\pi a\omega\mu}\frac{Q_\varphi}{Q}\right)^{1/2}. \tag{20}$$

The effective wave number $\tilde{k}$ defines the normalised resonance wavelength ($k_{res}l=\pi(2n-1)$):[19]

$$\lambda_{res,n}=\frac{2l}{2n-1}\sqrt{\varepsilon\mu}\,\mathrm{Re}\left(1+\frac{ic\varsigma_{xx}}{2\pi a\omega\mu}\frac{Q_\varphi}{Q}\right)^{1/2}. \tag{21}$$

$n=1,\ 2,\ 3\ \ldots$

Further we will consider only the first resonant frequency $f_{res}=c/\lambda_{res,1}$ at which the composite produces a maximal response.

## V. Field dependent impedance matrix

The field dependence of the effective permittivity $\varepsilon_{eff}$ of the composite is caused by the field dependence of the surface impedance matrix $\hat{\varsigma}(H_{ex})$, which determines the losses inside the inclusions. These internal losses characterise the quality factor of the entire composite system and the type of dispersion of $\varepsilon_{eff}$. The magneto-impedance matrix $\hat{\varsigma}(H_{ex})$ is a generalisation[23] of so-called magneto-impedance (MI) effect[5,6] in ferromagnetic materials. The calculation of $\hat{\varsigma}$ is based on the solution of the Maxwell's equations inside the conductor for the ac fields **e** and **h** together with the equation of motion for the magnetisation vector $\mathbf{M}$.[23] An analytical treatment is possible in a linear approximation with respect to the time-variable parameters $\mathbf{e}$, $\mathbf{h}$, $\mathbf{m} = \mathbf{M} - \mathbf{M}_0$, where $\mathbf{M}_0$ is the static magnetisation. Assuming a local relationship between $\mathbf{m}$ and $\mathbf{h}$: $\mathbf{m} = \hat{c}\,\mathbf{h}$, the problem is simplified to finding the solutions of the Maxwell equations with a given ac permeability matrix $\hat{\mu} = 1 + 4\pi\hat{\chi}$. In general, the anisotropy axis $\mathbf{n}_K$ has an angle $\psi$ with the wire axis (x-axis), as shown in Fig. 1. The wire is assumed to be in a single domain state with the static magnetisation $\mathbf{M_0}$ directed in a helical way having an angle $\theta$ with the x-axis. The external magnetic field $H_{ex}$ is assumed to be parallel to the wire axis. The field applied in the perpendicular direction does not effect the magnetic configuration. Then, in the system of randomly oriented wires the magnetic properties need to be averaged over the field orientation. The stable direction of $\mathbf{M_0}$ is found by minimising the magnetostatic energy *U*:[25]

$$\begin{gathered}\partial U / \partial \theta = 0,\\ U = -K\cos^2(\psi - \theta) - M_0 H_{ex}\cos(\theta),\end{gathered} \tag{22}$$

where *K* is the anisotropy constant. Equation (22) describes the rotational magnetisation process demonstrated in Fig. 2 where the magnetisation plots are given for three types of anisotropy: longitudinal ($\psi = 0$), circumferential ($\psi = 90°$) and helical ($\psi = 60°$).

The susceptibility matrix $\hat{c}$ has the following form in the co-ordinate system $(x, \varphi, r)$:[23]

$$\hat{\chi} = \begin{pmatrix} \chi_1 & -i\chi_a \cos(\theta) & i\chi_a \sin(\theta) \\ i\chi_a \cos(\theta) & \chi_2 \cos^2(\theta) & -\chi_2 \sin(\theta)\cos(\theta) \\ -i\chi_a \sin(\theta) & -\chi_2 \sin(\theta)\cos(\theta) & \chi_2 \sin^2(\theta) \end{pmatrix}, \tag{23}$$

where

$$\chi_1 = \omega_M (\omega_1 - i\,\tau\,\omega)/\Delta,$$
$$\chi_2 = \omega_M (\omega_2 - i\,\tau\omega)/\Delta,$$
$$\chi_a = \omega\omega_M /\Delta,$$
$$\Delta = (\omega_2 - i\,\tau\,\omega)(\omega_1 - i\tau\,\omega) - \omega^2,$$
$$\omega_1 = \gamma[H_{ex}\cos(\theta) + H_K \cos 2(\psi - \theta)],$$
$$H_K = 2K/M_0$$
$$\omega_2 = \gamma[H_{ex}\cos(\theta) + H_K \cos^2(\psi - \theta)],$$
$$\omega_M = \gamma M_0.$$

Here $g$ is the gyromagnetic constant, $\tau$ is the spin-relaxation parameter, $H_K$ is the anisotropy field. The impedance matrix $\hat{V}$ has been calculated in Ref. 23 for the strong and weak skin-effects. In this paper we give only the longitudinal impedance component $V_{xx}$ used for further calculations:

a) strong skin-effect ($\delta_m / a << 1$)

$$V_{xx} = \frac{c(1-i)\left(\sqrt{\tilde{m}}\cos^2(q) + \sin^2(q)\right)}{4p\ s\ d}, \tag{24}$$

b) weak skin-effect ($a/\delta << 1$ or $a/\ d \sim 1$, $a/\delta_m \sim 1$)

$$V_{xx} = \frac{k_1 c}{4p\ s}\frac{J_0(k_1 a)}{J_1(k_1 a)},$$

where $\tilde{m} = 1 + 4p\,\tilde{c}$ and $\tilde{c} = c_2 - 4p\ c_a^2/(1 + 4p\ c_1)$ are the effective permeability and susceptibility respectively, $\delta = c/\sqrt{2\pi\sigma\omega}$ is the nonmagnetic skin-depth, $J_{0,1}$ are the Bessel functions, $m_1 = 1 + 4p\cos^2(q)\,\tilde{c}$, $k_1^2 = m_1\left(4p\,jw\ s / c^2\right)$, $d_m = c/\sqrt{2p\ s\ w\ m_1}$ is the magnetic skin-depth.[23] Equations (24a) and (24b) demonstrate that the components of surface impedance matrix depend on both the ac susceptibility parameter $\tilde{c} = (\tilde{m} - 1)/4p$ and the static magnetisation orientation angle $\theta$. At high frequencies the latter will give the main contribution to the field dependence of the impedance since $\tilde{c}$ looses its the field sensitivity.

Usually ferromagnetic microwires have a circular ($\psi = 90^0$) or longitudinal ($\psi = 0$) magnetisation in the outer shell. The central part of the wire always consists of a longitudinally magnetised inner core. Wires with a circular anisotropy exhibit the most sensitive MI effect. Here, we use the parameters of Co-rich glass-coated wires with a negative magnetostriction and low coercitivity.[4,26] A circular "bamboo-like" domain structure[27] with opposite magnetizations in the adjacent domains exists almost in the entire wire which exhibits nearly a non-hysteretic B-H curve, as shown in Fig. 2 for $\psi = 90^0$. Due to such domain structure, the averaged off-diagonal components $\varsigma_{x\varphi}$ and $\varsigma_{\varphi x}$ are zero, as was proved theoretically[23] and experimentally.[28] Thus, the effects related to the off-diagonal components (see Eq. (16)) are possible only in a wire with a helical magnetisation where such averaging does not occur.

Figure (3) demonstrates the dispersion curves for the effective permeability parameter $\tilde{m} = 1 + 4p\,\tilde{c}$, which enters the impedance matrix in combination with the magnetisation angle $\theta$. The following magnetic parameters have been chosen: anisotropy field $H_K = 2$ Oe, saturation magnetisation $M_0 = 500$ G, giromagnetic constant $g = 2\times 10^7$ (rad/s)/Oe. In calculations, a small dispersion of the anisotropy angle $\psi$ with respect to $90^0$ should be introduced to model a real sample and avoid zero ferromagnetic resonance frequency at $H_{ex} = H_K$. The real part $\mathrm{Re}(\tilde{m}(w))$ approaches unity at the ferromagnetic resonance frequency ($\psi = 90°$) $f_{FMR} = \gamma\sqrt{|H_{ex} - H_K|4\pi M_0}/2\pi$: $f_{FMR} = 365$ MHz at $H_{ex} = 0$ and $f_{FMR} = 725$ MHz at $H_{ex} = 10$ Oe. In the gigahertz range, $\mathrm{Re}(\tilde{m}(w))$ is negative being in magnitude in the range of 10-20, and $\mathrm{Im}(\tilde{\mu}(\omega))$ is in the range of 10-40. Both of them become insensitive to the external magnetic field, as shown in Fig. 3(c). In this case, the field dependence of the impedance is entirely related with that for the static magnetisation orientation $\theta$ (see Eqs. (23),(24)). Then, if $\theta$ is a sensitive function of $H_{ex}$, to insure high field sensitivity of the impedance, it is important only that the condition $|\tilde{\mu}(\omega)| >> 1$ is held. This conclusion clearly demonstrates that the condition of the ferromagnetic resonance is not required for the MI effect, contrary to the widely expressed belief.[29,30]

A large difference between $f_{FMR}$ and the frequency where the imaginary part reaches a maximum value is caused by the specific form of the effective susceptibility $\tilde{\chi}$ containing all the components of the susceptibility matrix $\hat{\chi}$ :[23]

$$\tilde{\chi} = \frac{\omega_M(\omega_2 - i\tau\omega) + 4\pi\omega_M^2}{(\omega_1 - i\tau\omega)(\omega_2 + 4\pi\omega_M - i\tau\omega) - \omega^2} \qquad (25)$$

The expression for $f_{FMR}$ directly follows from Eq. (25): $f_{FMR} = \sqrt{\omega_1(\omega_2 + 4\pi\omega_M)}/2\pi$. The dispersion curves, considered above, look very similar to a relaxation spectrum typical of polycrystalline multidomain ferrites. However, in our case, the “relaxation-like” dispersion is caused by a complicated form (25) of the effective susceptibility. Such kind of the dispersion is always observed in experiments with bulk ferromagnetic conductive samples, where the skin-effect is important and the effective susceptibility is composed of the components of the internal matrix $\hat{\chi}$.[31]

Figures 4 demonstrate the field dependence of $\varsigma_{xx}(H_{ex})$ for the GHz range. The wire has 10 $\mu m$ diameter, conductivity $\sigma = 7.6\times10^{15}$ $s^{-1}$, anisotropy field $H_K = 2$ Oe, saturation magnetisation $M_0 = 500$ G, and giromagnetic constant $\gamma = 2\times10^{7}$ (rad/s)/Oe. The calculations have been carried out in a low frequency limit (see Eq. (24b)) since the magnetic skin-depth $\delta_m = c/\sqrt{2\pi\sigma\omega\mu_1}$ is of the order of the wire radius. As it will be shown below, not very strong skin effect is the basic requirement for obtaining the field-dependent effective permittivity. For this reason the wire diameter is chosen to be sufficiently small. At the high frequencies the curve of $\varsigma_{xx}(H_{ex})$ flattens at $H_{ex} > H_K$ and even for large values of $H_{ex} >> H_K$ does not reach the saturation. In the gigahertz range $\varsigma_{xx}(H_{ex})$ has approximately a constant value for $H_{ex} \geq H_K$, as shown in Fig. 4, reflecting the field dependence of $\cos^2(\theta)$ (see Eqs. (24a,b) and Fig. 2) since the permeability parameter $\tilde{\mu}$ looses its field sensitivity (as discussed above in Fig. 3(c)). Such kind of the transformation of $\varsigma_{xx}(H_{ex})$ is supported by a number of experiments where the magneto-impedance spectra were measured in ferromagnetic wires in a wide frequency range from 1 kHz up to 1.2 GHz.[29,32] The field dependence $\varsigma_{xx}(H_{ex})$ at very high frequencies may be of a considerable interest since it has two predetermined levels at $H_{ex} = 0$ and $H_{ex} > H_K$, between which a smooth transition can be achieved by applying $H_{ex}$. Thus, at very high frequencies the impedance $\varsigma_{xx}(H_{ex}) \sim \cos^2(\theta)$ exhibits a "valve-like" behaviour, switching from one stable level to the other, following the dc magnetisation.

**VI. Field dependent resonance effective permittivity**

It is quite natural to expect that the field dependence $\varepsilon_{eff}(H_{ex})$ becomes most sensitive in the vicinity of the antenna resonance where any small variations in the inclusion parameters cause a strong change of the current distribution and the inclusion dipole moment. This results in a remarkable transformation of the dispersion curve $\varepsilon_{eff}(\omega)$ under the external magnetic field. The resonance frequency range is determined by the wire length $l$ and the matrix permittivity $\varepsilon$. Practically, it is not desirable to construct composite materials with inclusions having a length larger than 1 cm. In this case, the first resonance frequency in air $f_{res} = c/2l$ would be in the range of tens gigahertz. However, for such high frequencies the magnetic properties of soft ferromagnetic wires tail off completely and $\tilde{m}$ tends to be unity. Without increasing the wire length, the operating frequencies can be lowered (in $\sqrt{\varepsilon}$ times) by using a dielectric matrix with higher permittivity $\varepsilon >> 1$: $f_{res} = c/2l\sqrt{e}$ .

The condition of a moderate skin-effect ($a/\delta_m \sim 1$) is proving to be important to realise a high sensitivity of $\varepsilon_{eff}(H_{ex})$. If the magnetic skin-depth is much smaller than the wire radius ($\delta_m / a << 1$), the normalised wave number $\tilde{k}$ differs little from the wave number $k$ of the free space. Substituting the high frequency impedance (24a) into (20) gives:

$$\tilde{k} \sim \frac{w}{c}\sqrt{em}\left(1 + \frac{(1-i)}{2m\, l \ln(l/a)} \frac{d}{a} \sqrt{\tilde{m}} \cos^2(q)\right)^{1/2} \tag{26}$$

From Eq. (26) it immediately follows that if $\delta / a << 1$ the wave number becomes $\tilde{k} \approx \omega\sqrt{\varepsilon\mu}/c$, that is an essential field dependence $\varepsilon_{eff}(H_{ex})$ can be reached in the case when the nonmagnetic skin depth is of the order of wire radius.

At a very low inclusion concentration $p << p_c \propto 2a/l$ the effective permittivity $\varepsilon_{eff}(\omega)$ can be represented by Eq. (3) as the dipole sum with the polarisability $<\alpha>$ averaged over the inclusion orientations. The polarisability $a$ has to be calculated from Eq. (1) using the first $j_1(x)$ approximation for the current distribution, which takes into account all the losses in the system (see Eq. (A9) in Appendix). In the case of a planar composite shown in Fig. 5, the averaging gives a coefficient 1/2 ($<\alpha>=\alpha/2$) since the ac electric field perpendicular to the wire does not cause significant polarisation effects. Then $<\alpha>$ determines the effective permittivity in the plane of the sample. The total permittivity matrix of a planar composite is of the form:

$$\hat{\varepsilon}(H_{ex}) = \begin{pmatrix} \varepsilon_{eff}(H_{ex}) & 0 & 0 \\ 0 & \varepsilon_{eff}(H_{ex}) & 0 \\ 0 & 0 & \varepsilon \end{pmatrix} \quad (27)$$

Figure 6 demonstrates the dispersion curves of the effective permittivity $\varepsilon_{eff}(\omega)$ at the gigahertz range as a function of the external magnetic field $H_{ex}$ for $\varepsilon = 16$ and volume concentration $p = 0.01\%$. This concentration is considerably smaller than the percolation threshold $p_c \propto (2a/l) \times 100\% \sim 0.1\%$ ($2a = 10\,\mu m$ and $l = 1\,cm$). The field-dependence effect shows up in changing the character of the dispersion curves. In the absence of $H_{ex}$ the dispersion curves are of a resonance type: at $f = f_{res}$ the imaginary part reaches a maximum and the real part equals zero. Applying a magnetic field $H_{ex} > H_K$, the impedance $\varsigma_{xx}(H_{ex})$ is increased and, as a consequence, the internal losses in the inclusion, which results in the dispersion of a relaxation type. In the presence of $H_{ex}$ the resonance frequency also slightly shifts towards higher frequencies. In Ref. 10 the transformation of the dispersion $\varepsilon_{eff}(\omega)$ from a resonant type to relaxation one was associated with a different wire conductivity $\sigma$, which defines the resistive loss. In our case it is provided through the field dependent impedance $\varsigma_{xx}(H_{ex})$ instead of conductivity. At the wire concentration larger than a certain value (for example, $p = 0.01\%$) the real part of $\varepsilon_{eff}(\omega)$ becomes negative in the vicinity of the resonance. Applying $H_{ex}$, it is possible change gradually $\mathrm{Re}(\varepsilon_{eff}(\omega))$ from negative to positive values.

The field-dependent permittivity matrix $\hat{\varepsilon}(H_{ex})$ can be used in tuneable microwave covers. Figure 7 shows the dispersion curves of the reflection $R(\omega)$ and transmission $T(\omega)$ coefficients for a thin wire-composite sheet in the case of normal incidence of the electromagnetic wave for two values of $H_{ex}$. The magnetic field can be applied in any direction (see Fig. 5). The energy absorption A=(1-R-T) in such kind of a composite is rather strong, therefore it has to be sufficiently thin to be used as a wave passage. Another promising application is suggested to employ these composites as an internal cover in the partially filled waveguides or layered waveguides with a “dielectric/composite” structure. The waveguide with an internal composite cover is proving to operate similar to the waveguide containing absorption ferromagnetic layers,[33] for example, a thin-film “dielectric/Fe” structure considered in Ref. 34,35. In both systems there is an anisotropic field-dependent layer with resonance properties and large energy losses, but in our case the field-dependent layer is made of a composite material with $\hat{e}(H_{ex})$. The operating frequency range is determined by the antenna resonance. Such a waveguide system, by analogy to that already designed for the waveguides with ferromagnets, can be used for tuneable filters and phase shifters operating up to tens gigahertz.

**Conclusion**

A comprehensive analysis of magnetic-field dependent dielectric response in diluted metal-dielectric composites containing ferromagnetic microwires is presented. We have developed a rigorous mathematical method of calculating the electric current distribution at the wire-inclusion irradiated by an electromagnetic field, which determines the electric polarisability contributing to the effective permittivity. The wire polarisability is proven to be very sensitive to the surface impedance changes near the antenna resonance. Therefore, in the composites with ferromagnetic wires the effective permittivity may depend on a static magnetic field via the corresponding dependence of the impedance (known as magneto-impedance effect). The field dependence of the impedance remains substantial even for gigahertz frequencies much higher than the characteristic frequency of the ferromagnetic resonance. In this case, the impedance vs. field shows a “valve-type” behaviour reflecting the dc magnetisation reversal. In the case of a Co-based microwires, a moderate axial magnetic field of a few Oe magnetises the wire in the axial direction. During this process, the dispersion characteristics of permittivity can be changed considerably, say, from a resonance to a relaxation type. A number of applications of this effect is proposed including microwave materials with the field-dependent reflection/transmission coefficients and tuneable waveguides where the composite material is used as an additional field-dependent cover.

The theory can be generalised to the case of interacting wires and applied for the exact calculation of the effective response from composites containing periodically spaced wires (electromagnetic crystals). This presents a considerable interest for studding field-tuneable band-gap[7] and negative index materials,[11-13] and is intended to be published elsewhere.

**Appendix**

The iteration procedure proposed here allows the analytical expression for the current density $j(x)$ with the account of all the losses in the system to be obtained. Let us write once again the basic integro-differential equation (16):

$$\frac{\partial^2}{\partial x^2}(G*j)+\left(\frac{\omega}{c}\right)^2\varepsilon\mu\left[(G*j)+\frac{ic\varsigma_{xx}}{2\pi a\omega\mu}(G_\varphi*j)\right]=\frac{i\omega\varepsilon}{4\pi}\left[\bar{e}_{0x}(x)+\varsigma_{x\varphi}\bar{h}_{0x}(x)\right] \qquad (A1)$$

The general solution of Eq. (A1) for the zero approximation (19) has the following form:

$$j_0(x)=A\sin(\tilde{k}\,x)+B\cos(\tilde{k}\,x)+\frac{i\omega\varepsilon}{4\pi Q\tilde{k}}\int_{-l/2}^{x}\sin(\tilde{k}(x-s))(\bar{e}_{0x}(s)+\varsigma_{x\varphi}\bar{h}_{0x}(s))ds\,. \qquad (A2)$$

For the zero approximation with $\bar{e}_{0x}=\bar{e}_0=const$ and $\bar{h}_{0x}=\bar{h}_0=const$ we obtain:

$$j_0(x)=A\sin(\tilde{k}\,x)+B\cos(\tilde{k}\,x)+\frac{i\omega\varepsilon(\bar{e}_0+\varsigma_{x\varphi}\bar{h}_0)}{4\pi Q\tilde{k}^2}\,. \qquad (A3)$$

Zero approximation (A3), satisfying the boundary conditions $j_0(-l/2)=j_0(l/2)\equiv 0$, is of the form:

$$j_0(x)=\frac{i\omega\varepsilon(\bar{e}_0+\varsigma_{x\varphi}\bar{h}_0)}{4\pi Q\tilde{k}^2}\frac{(\cos(\tilde{k}\,l/2)-\cos(\tilde{k}\,x))}{\cos(\tilde{k}\,l/2)} \qquad (A4)$$

For the ideal conductor ($\sigma=\infty$) and $\mathrm{Im}(\varepsilon)=\mathrm{Im}(\mu)\equiv 0$ solution (A4) has singularities at the resonance wavelengths $\lambda_{res}$ defined by Eq. (21). The account of a limited conductivity and radiation losses eliminates the singularities.

The iteration procedure determines the next approximations. Let us extract the real and imaginary parts of the convolutions in Eq. (A1). To calculate the convolutions with real parts the proposed method given by Eq. (18). As a result we obtain the following integro-differential equation:

$$\frac{\partial^2}{\partial x^2}\left[ j(x)+\frac{i}{Q}(\mathrm{Im}(G)*j)\right]+\tilde{k}^2\left[ j(x)+\frac{i}{Q}(\mathrm{Im}(G)*j)\right]=$$
$$=\frac{i\omega\varepsilon}{4\pi Q}\left[\bar{e}_{0x}(x)+\varsigma_{x\varphi}\bar{h}_{0x}(x)\right]+\frac{i(\tilde{k}^2-k^2)}{Q}(\mathrm{Im}(G)*j)+\frac{\omega\varepsilon\,\varsigma_{xx}}{2\pi acQ}(\mathrm{Im}(G_\varphi)*j) \quad . \qquad \text{(A5)}$$

Equation (A5) can be considered as a non-uniform differential equation with respect to the operator $\partial^2/\partial x^2+\tilde{k}^2$. Finding the inverse operator, we obtain the integral equation:

$$j(x)=A\sin(\tilde{k}x)+B\cos(\tilde{k}x)+\frac{i\omega\varepsilon}{4\pi Q\tilde{k}}\int_{-l/2}^{x}\sin(\tilde{k}(x-s))(\bar{e}_{0x}(s)+\varsigma_{x\varphi}\bar{h}_{0x}(s))ds+$$
$$+\frac{i(\tilde{k}^2-k^2)}{Q\tilde{k}}\int_{-l/2}^{x}\sin(\tilde{k}(x-s))(\mathrm{Im}(G)*j))ds+ \quad , \qquad \text{(A6)}$$
$$+\frac{\omega\varepsilon\,\varsigma_{xx}}{2\pi acQ\tilde{k}}\int_{-l/2}^{x}\sin(\tilde{k}(x-s))(\mathrm{Im}(G_\varphi)*j)ds-\frac{i}{Q}(\mathrm{Im}(G)*j)$$

where the constants $A$ and $B$ should be chosen to satisfy the boundary conditions $j(-l/2)=j(l/2)\equiv 0$. Equation (A6) is the Fredholm integral equation of the second kind, therefore it is well adapted to an iterative method. For the n-th estimation the following iteration procedure is used:

$$j_n(x)=j_0(x)+\frac{i(\tilde{k}^2-k^2)}{Q\tilde{k}}\int_{-l/2}^{x}\sin(\tilde{k}(x-s))(\mathrm{Im}(G)*j_{n-1}))ds+$$
$$+\frac{\omega\varepsilon\,\varsigma_{xx}}{2\pi acQ\tilde{k}}\int_{-l/2}^{x}\sin(\tilde{k}(x-s))(\mathrm{Im}(G_\varphi)*j_{n-1})ds-\frac{i}{Q}(\mathrm{Im}(G)*j_{n-1}) \quad . \qquad \text{(A7)}$$

Constants $A$ and $B$ are calculated at the final step of the iteration procedure $n=N>1$ so that to satisfy the boundary conditions $j_N(-l/2)=j_N(l/2)\equiv 0$.

Let us rewrite iteration representation (A7) in the following compact form:

$$j_n(x)=j_0(x)+\int_{-l/2}^{l/2}S(x,q)\,j_{n-1}(q)dq\,, \qquad \text{(A8)}$$

where

$$S(x,q) = S_1(x,q) + S_2(x,q) + S_3(x,q),$$

$$S_1(x,q) = -\frac{i}{Q}\,\mathrm{Im}(G(r)), \quad r = \sqrt{(x-q)^2 + a^2},$$

$$S_2(x,q) = \frac{i(\tilde{k}^2 - k^2)}{Q\tilde{k}} \int_{-l/2}^{x} \sin(\tilde{k}(x-s))\,\mathrm{Im}(G(r))ds, \; r = \sqrt{(s-q)^2 + a^2},$$

$$S_3(x,q) = \frac{\omega\varepsilon\,\varsigma_{xx}}{2\pi a c Q\tilde{k}} \int_{-l/2}^{x} \sin(\tilde{k}(x-s))\,\mathrm{Im}(G_\varphi(r))ds, \; r = \sqrt{(s-q)^2 + a^2}$$

The first iteration will be enough to take into account the radiation losses. Using Eq. (A3) as the zero approximation we obtain:

$$\begin{aligned} j_1(x) = A\left(\sin(\tilde{k}\,x) + \int_{-l/2}^{l/2} S(x,q)\sin(\tilde{k}\,q)dq\right) + B\left(\cos(\tilde{k}\,x) + \int_{-l/2}^{l/2} S(x,q)\cos(\tilde{k}\,q)dq\right) + \\ + \frac{i\,\omega\varepsilon(\bar{e}_0 + \varsigma_{x\varphi}\bar{h}_0)}{4\pi Q\tilde{k}^2}\left(1 + \int_{-l/2}^{l/2} S(x,q)dq\right) \end{aligned} \quad \text{(A9)}$$

The unknown constants $A$ and $B$ are to be found from the boundary conditions $j_1(-l/2) = j_1(l/2) \equiv 0$.

A detailed analysis of the radiation effects for antennas of different shapes (including nanoantennas) is considered in Ref. 36.

## Figures:

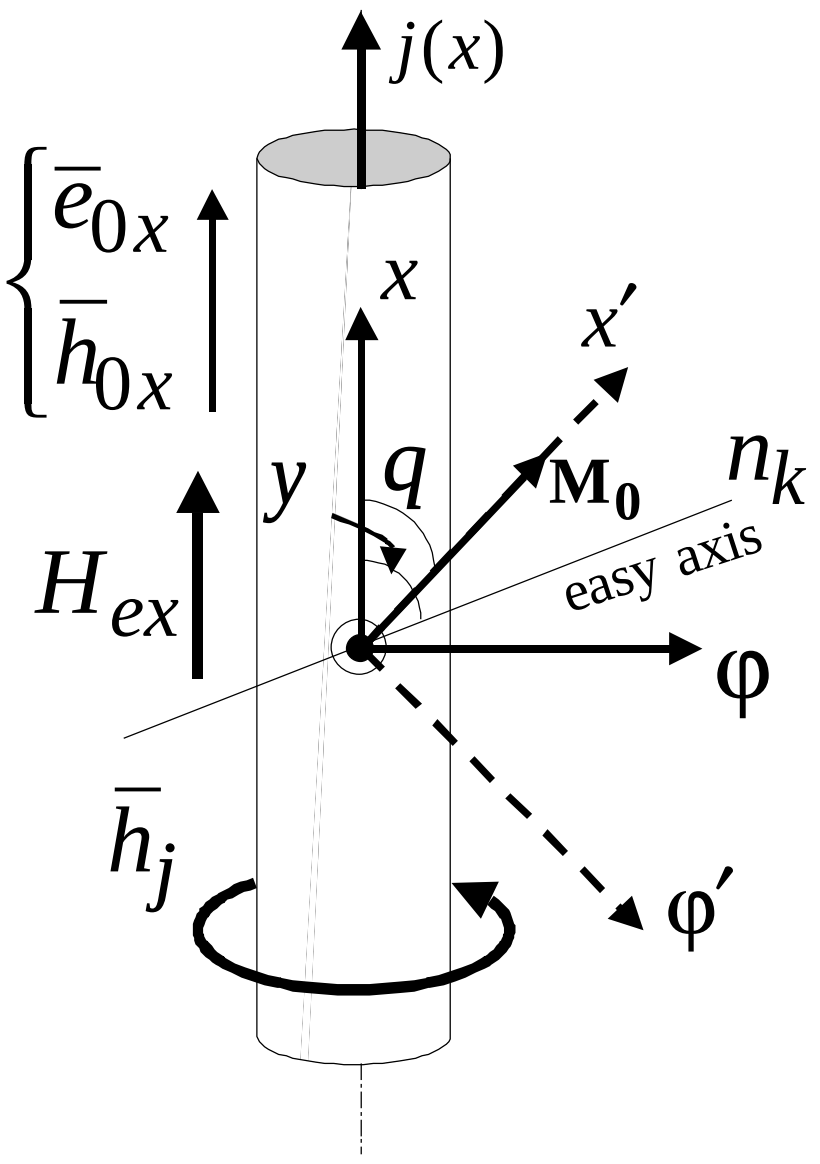

**Fig. 1.** Schematic diagram of the magnetic configuration in a wire.

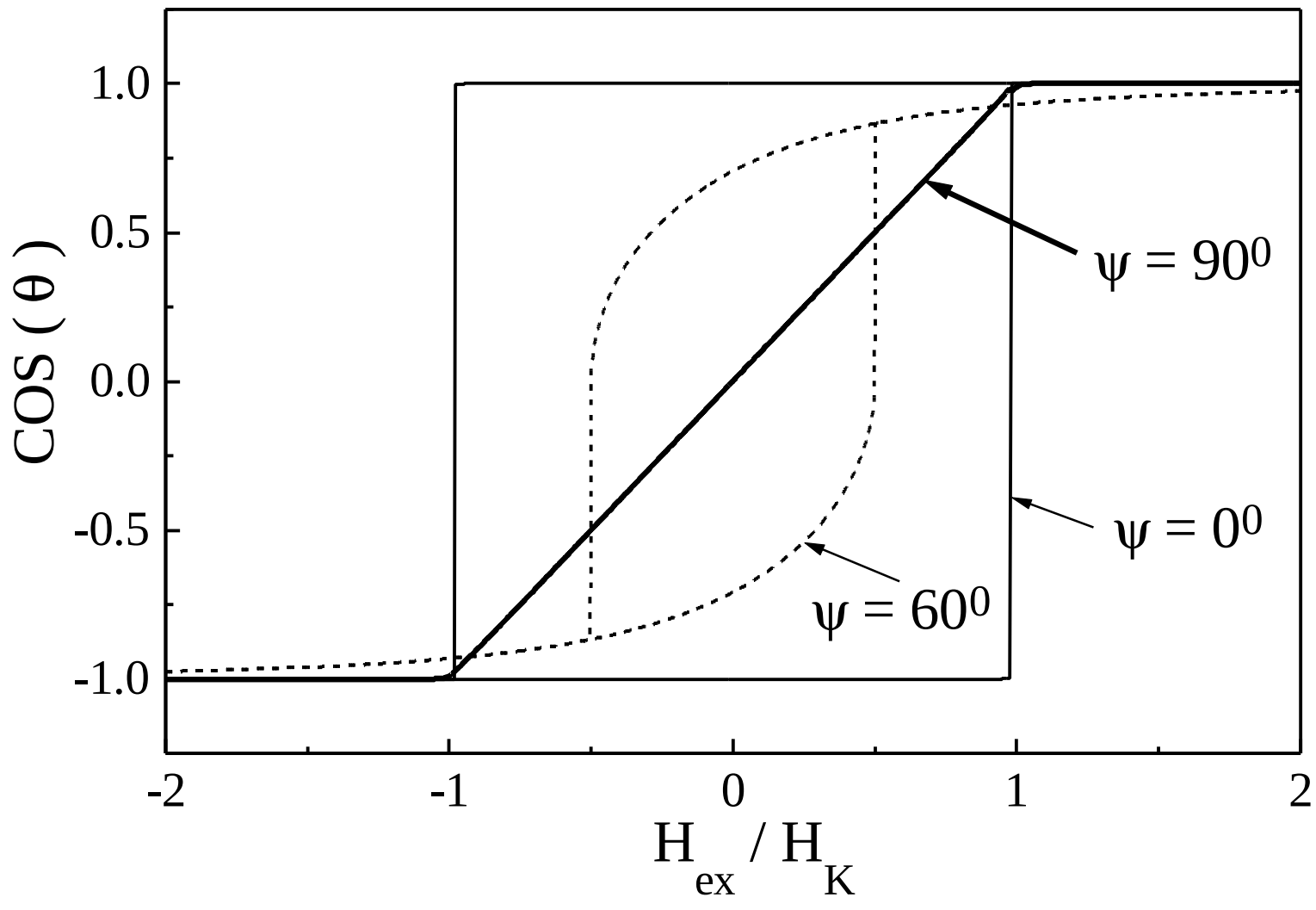

**Fig. 2.** Typical rotational hysteresis curves $M_{0x}(H_{ex})$ for different types of anisotropy: longitudinal ( $\psi = 0$ ), circumferential ( $\psi = 90^0$ ), and helical ( $\psi = 60^0$ ).

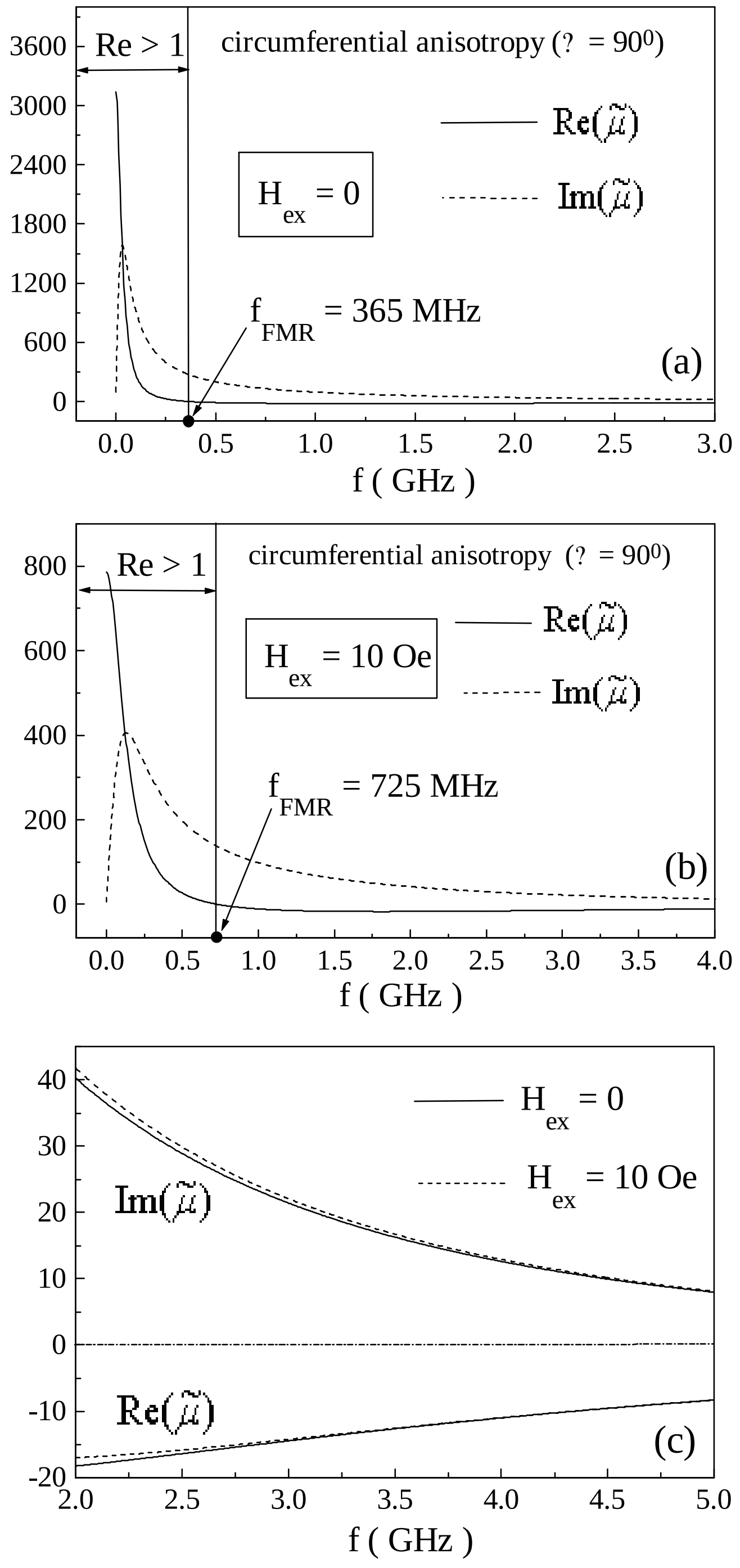

**Fig. 3.** Dispersion curves of the effective permeability $\tilde{m}$ for different $H_{ex}$.

Magnetic parameters: anisotropy field $H_K = 2$ Oe, saturation magnetisation $M_0 = 500$ G, and giromagnetic constant $g = 2\times10^7$ (rad/s)/Oe. For frequencies much higher than $f_{FMR}$ (gigahertz range) $\tilde{m}$ becomes insensitive to $H_{ex}$ as shown in (c).

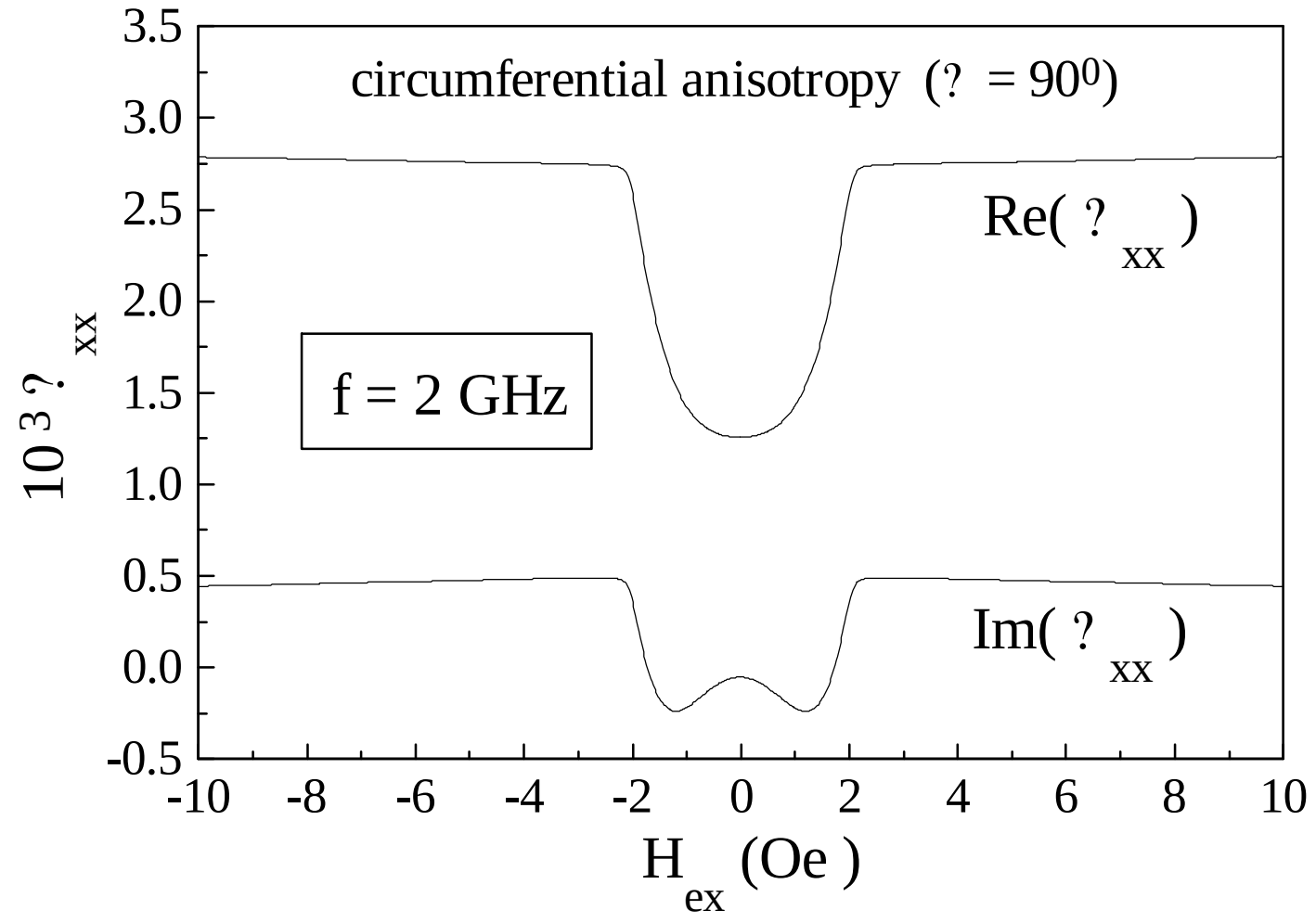

**Fig. 4.** Typical "valve-like" field dependence of the longitudinal impedance in the gigahertz range: $\varsigma_{xx}(H_{ex})$ is approximately constant for $H_{ex} \geq H_K$ reflecting the field dependence of $\cos^2(\theta)$ since $\tilde{m}$ looses its field sensitivity (as shown in Fig. 3(c)).

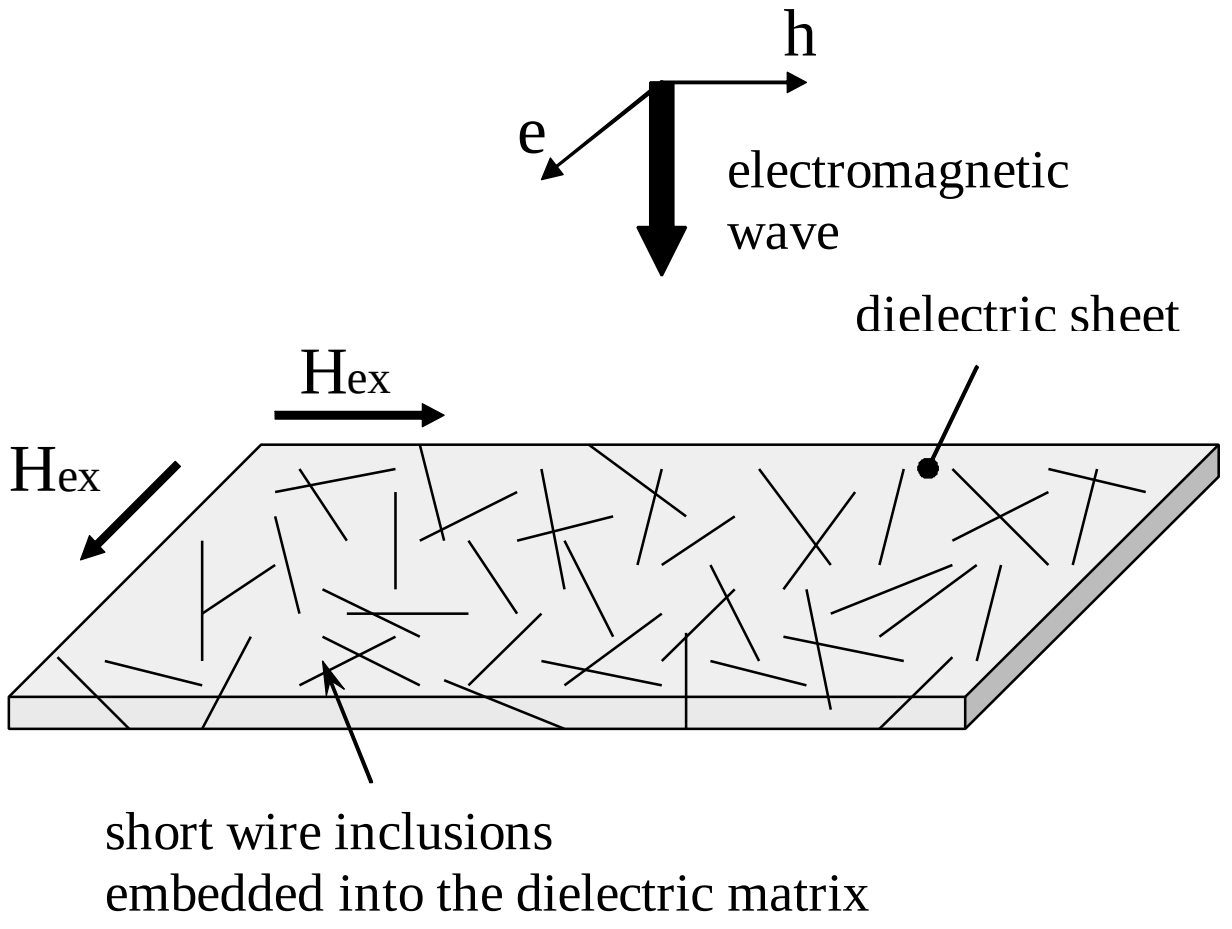

**Fig. 5.** Microstructure of a thin composite sample containing short wires. The sheet thickness $h$ is much smaller than the wire length $l$. The wire inclusions have random orientations in the plane of the sample. The magnetic field $H_{ex}$ can be applied in any direction.

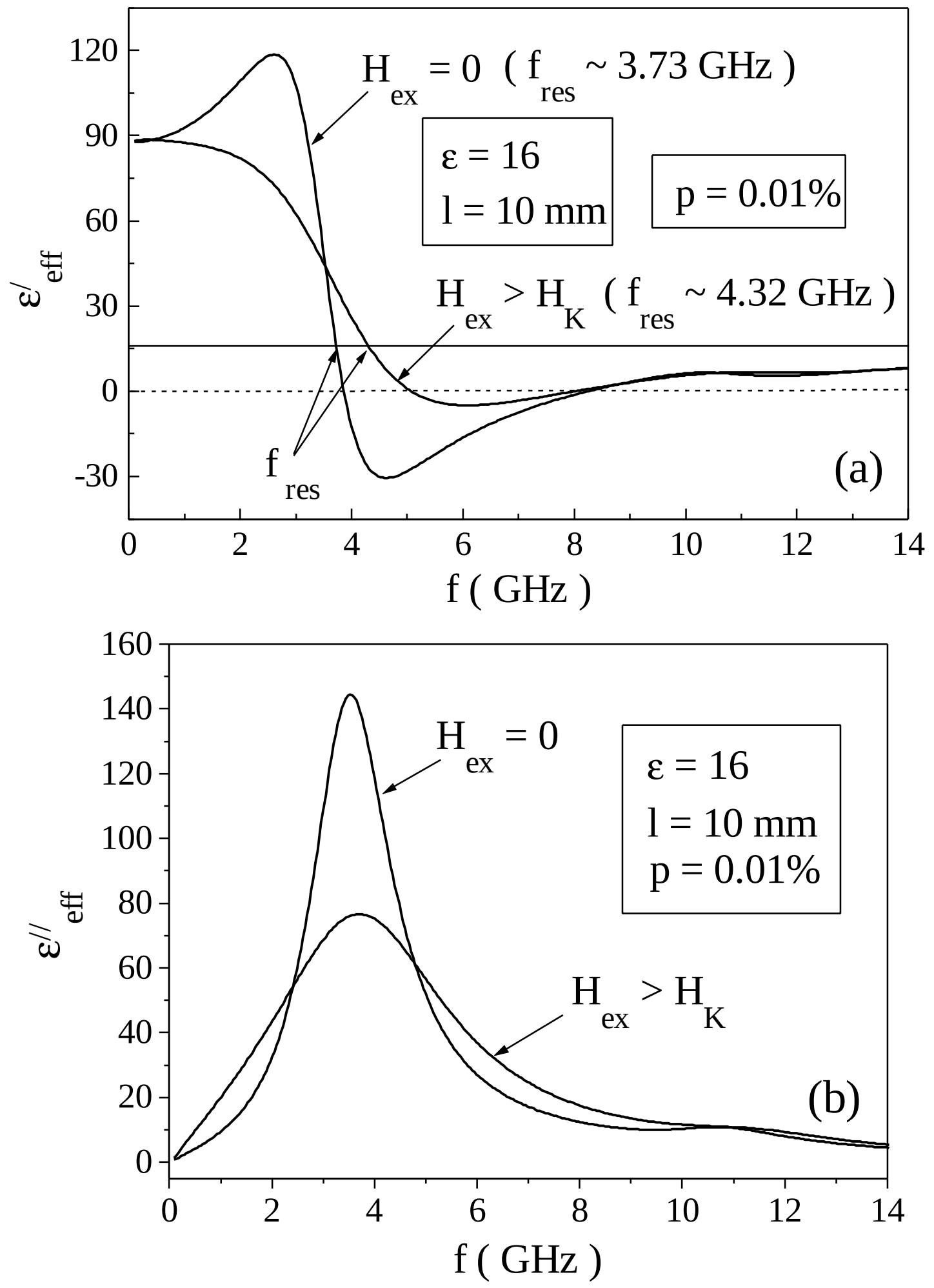

**Fig. 6.** Transformation of the dispersion of the effective permittivity from a resonance type to a relaxation one due to $H_{ex}$ in the vicinity of the antenna resonance. The inclusion concentration is $p = 0.01\%$ and matrix permittivity is $\varepsilon = 16$. Applying $H_{ex}$, the negative peak of $\mathrm{Re}(\varepsilon_{eff})$ continuously decreases as the dispersion tends to become of a relaxation type with $\mathrm{Re}(\varepsilon_{eff})>0$.

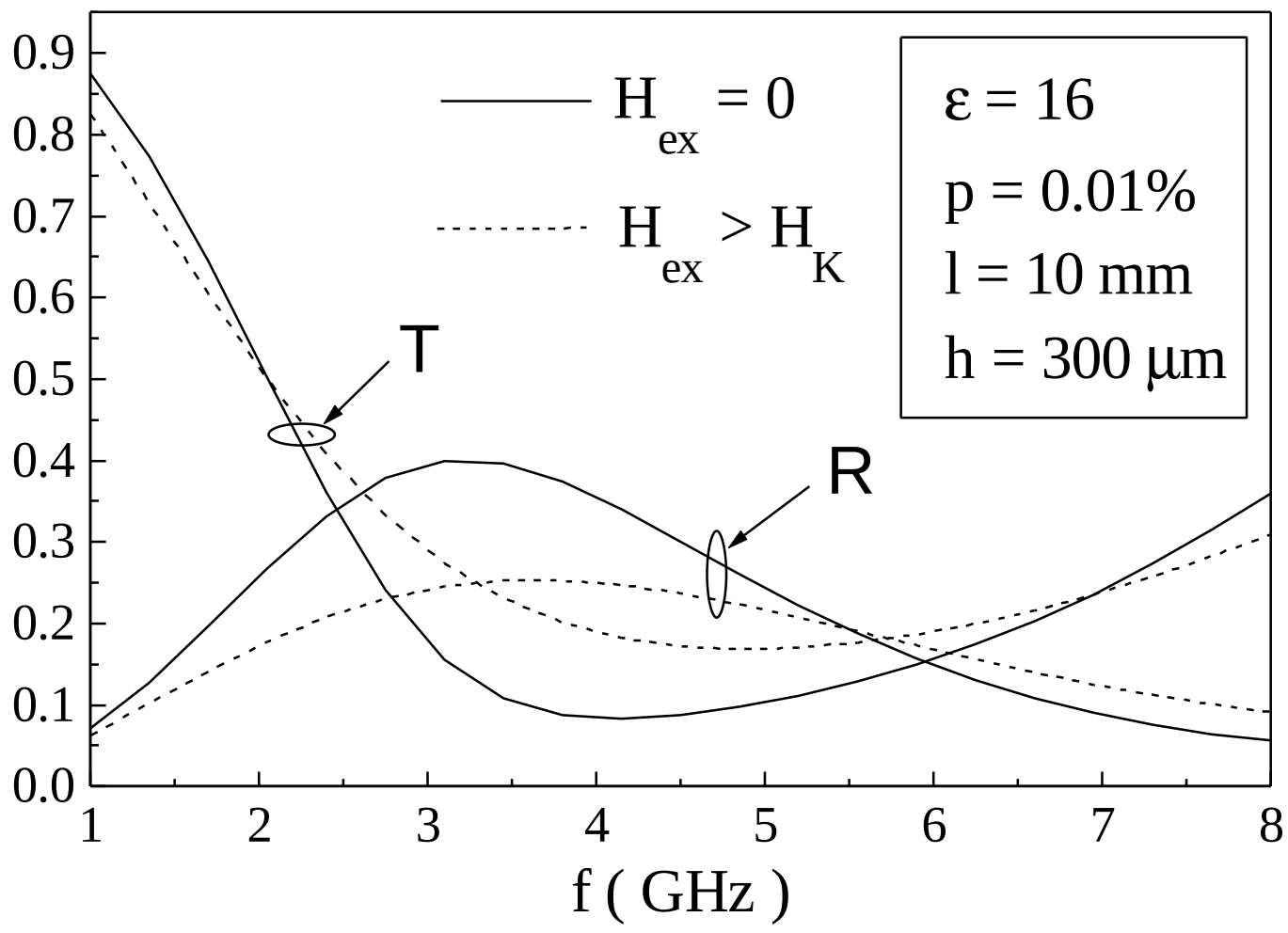

Fig. 7. Dispersion curves of the reflection $\mathbf{R}(\omega)$ and transmission $\mathbf{T}(\omega)$ coefficients for a thin wire-composite slab in the case of normal incidence of the electromagnetic wave as a function of $H_{ex}$. The sample thickness is $h = 300\mu m$.